\documentclass[conference]{IEEEtran}
\IEEEoverridecommandlockouts
\usepackage{cite}
\usepackage{amsmath,amssymb,amsfonts}
\usepackage{algorithmic}
\usepackage{graphicx}
\usepackage{subcaption}
\usepackage{textcomp}
\usepackage[table]{xcolor}
\usepackage{braket}
\usepackage{array}
\usepackage{makecell}
\usepackage{graphicx}
\usepackage{booktabs}
\usepackage{listings}
\usepackage{xparse}
\usepackage{comment}
\usepackage{balance}

\newcommand{\etal}{\emph{et al.}}
\newcommand{\ie}{\emph{i.e.}}

\newcommand{\eg}{\emph{e.g.}}

\newcolumntype{s}{>{\columncolor{blue!30} \bf} c}
\newcolumntype{b}{>{\bf} c}
\def\BibTeX{{\rm B\kern-.05em{\sc i\kern-.025em b}\kern-.08em
    T\kern-.1667em\lower.7ex\hbox{E}\kern-.125emX}}

\begin{document}

\title{Quantum Machine Learning for Anomaly Detection in Consumer Electronics\\
}

\author{\IEEEauthorblockN{Sounak Bhowmik}
\IEEEauthorblockA{\textit{Dept. of Electrical Engineering and Computer Science} \\
\textit{University of Tennessee, Knoxville}\\
Tennessee, USA \\
sbhowmi2@vols.utk.edu}
\and
\IEEEauthorblockN{Himanshu Thapliyal}
\IEEEauthorblockA{\textit{Dept. of Electrical Engineering and Computer Science} \\
\textit{University of Tennessee, Knoxville}\\
Tennessee, USA \\
hthapliyal@utk.edu}
}

\maketitle

\begin{abstract}
Anomaly detection is a crucial task in cyber security. Technological advancement brings new cyber-physical threats like network intrusion, financial fraud, identity theft, and property invasion. In the rapidly changing world, with frequently emerging new types of anomalies, classical machine learning models are insufficient to prevent all the threats. Quantum Machine Learning (QML) is emerging as a powerful computational tool that can detect anomalies more efficiently. In this work, we have introduced QML and its applications for anomaly detection in consumer electronics. We have shown a generic framework for applying QML algorithms in anomaly detection tasks. We have also briefly discussed popular supervised, unsupervised, and reinforcement learning-based QML algorithms and included five case studies of recent works to show their applications in anomaly detection in the consumer electronics field.
\end{abstract}


\section{Introduction}
Anomaly detection in consumer electronics refers to identifying irregular patterns that deviate from the normal functioning of devices we use daily. These anomalies can range from minor software glitches to significant security vulnerabilities. Consumer electronics, particularly IoT (Internet of Things) devices, are integral to our everyday lives, yet they are susceptible to various disruptions and cyber-attacks. For instance, an irregularity in a smart home system might compromise the security of an entire household. Just as anomalies in network traffic could signal security threats and unexpected patterns in medical scans might indicate health issues, irregularities in consumer electronics can have profound implications, from data breaches to complete system failures. Anomalies are everywhere, from credit card transactions to space-craft data and thermal power stations to our daily email services. Hence, to safeguard against these threats, Anomaly Detection Systems (ADS) is crucial and widely implemented across numerous sectors within the consumer electronics industry~\cite{10398426}, such as financial sectors (\eg, fraud detection), medical (\eg, disease diagnosis), surveillance (\eg, theft, robbery, property invasion), and cyber security (\eg, malware and network intrusion). Over the years, researchers have used ideas from statistics, machine learning, information theory, and spectral analysis to solve anomaly detection problems. Classical machine learning algorithms like clustering, one-class Support Vector Machines (SVMs), Decision Trees, and Neural Networks have been used successfully to build ADS. However, these algorithms are very resource-intensive and take a long time to train. They also suffer from over-fitting problems and find it challenging to adapt to new anomalies that could be introduced during their functional operation. Therefore, to find a better solution, many researchers are inclined towards a new discipline, Quantum Machine Learning (QML), that combines the power of quantum computing, quantum information, and machine learning algorithms. 

Quantum machine learning has been deployed for consumer electronics applications such as credit card fraud detection, anomaly detection in surveillance, health care anomaly detection, and crime prevention. Though literature surveys on anomaly detection using QML algorithms exist~\cite{maheshwari2022quantum}, there is no such comprehensive survey considering the anomalies in consumer electronic devices, to our knowledge. Therefore, this paper aims to introduce the generic framework of QML and its applications in anomaly detection in consumer electronics. The contributions of this work are as follows:
\begin{itemize}
    \item We introduced the generic framework of the QML algorithms used for anomaly detection in the consumer electronics domain.
    \item We have discussed various emerging QML algorithms, \eg, quantum neural network (QNN), quantum support vector machine (QSVM), hybrid QNN, quantum clustering, quantum ensemble methods, and quantum reinforcement learning, that can be used for anomaly detection in consumer electronics.
    \item To illustrate the real-world applications of QML for anomaly detection in consumer electronics, we have selected five case studies on the Internet of Medical Things (IoMT), surveillance, image processing, sensor intrusion, and credit card fraud detection to illustrate the utility of QML in anomaly detection in some consumer electronics fields.
    \item We conclude this paper by discussing new opportunities and challenges in this fast-growing discipline of QML in anomaly detection in consumer electronics.
\end{itemize}
\begin{figure*}[htb]
    \centering
    \includegraphics[width=2.1\columnwidth]{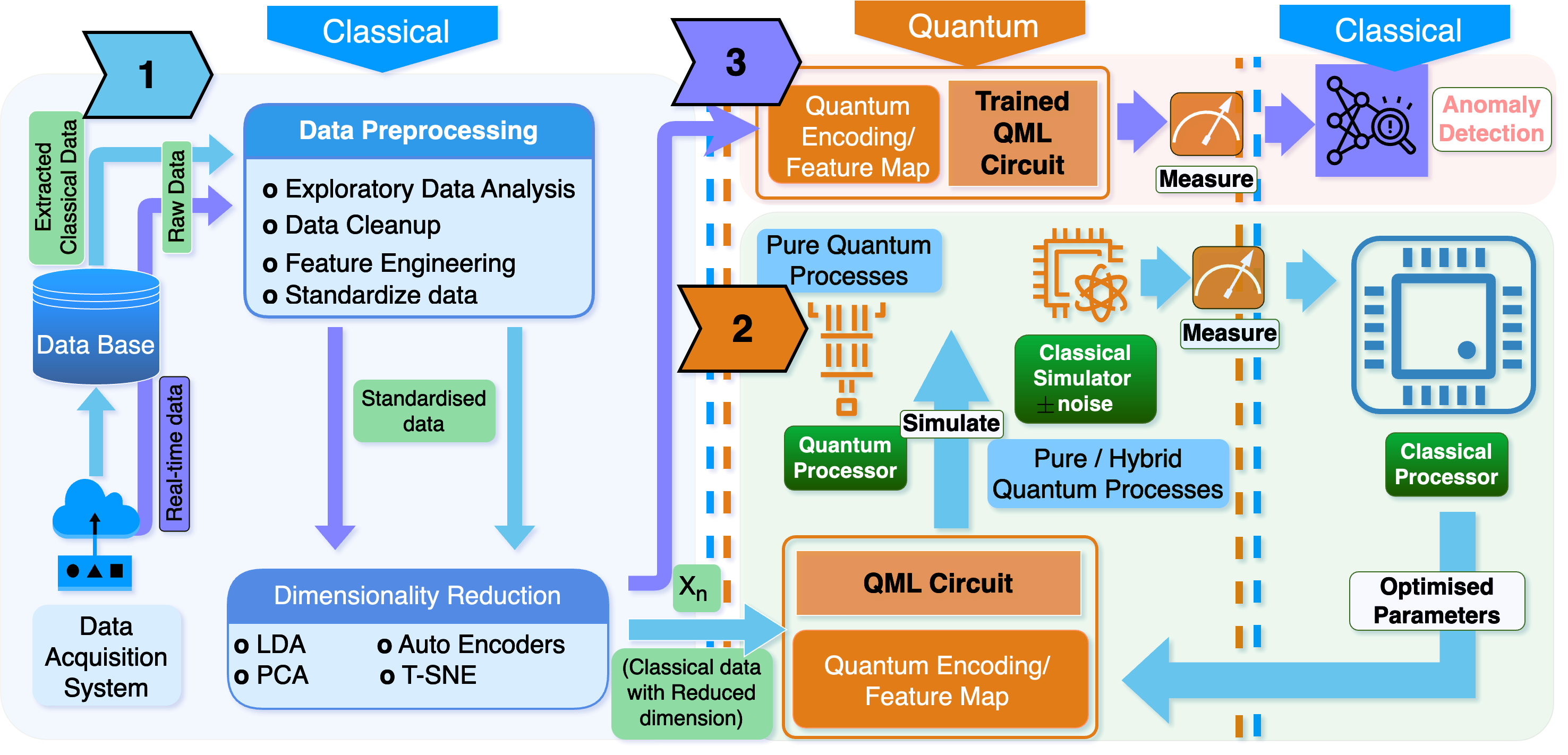}
    \caption{General Framework for Anomaly detection using} Quantum Machine Learning using Classical Data. There are three significant steps to building a QML-based anomaly detection system. The 1st step is to clean up the classical data and reduce the dimension as required. In the 2nd step, we feed clean data into the QML circuit after quantum encoding and optimize the model with a classical computer. After we have trained our model, we can feed data into it to get a prediction if the data is an anomaly in the 3rd step.
    \label{fig:QML_framework}
\end{figure*}

\section{Background On Quantum Machine Learning algorithms used in Anomaly Detection} \label{background}
Machine learning (ML) is a branch of statistics used for pattern recognition, feature extraction, classification, and generation, which solves many real-life problems, such as anomaly detection. ML algorithms refer to complex numerical models or functions with several learnable parameters. While training on a dataset, these parameters are optimized over multiple iterations, and the model learns to infer from unknown data. With a surge in total data storage globally, researchers are inclined to optimize the machine learning methods using the power of quantum computers under the umbrella of quantum machine learning. The fundamental unit of computation in a quantum computer is a qubit. A qubit is a mathematical object with states $\ket{0}$, $\ket{1}$ as well as $\alpha\ket{0} + \beta\ket{1}$, a superposition state. $\alpha$ and $\beta$ are complex numbers, representing the probability of the qubit collapsing to either $\ket{0}$ or $\ket{1}$ upon measurement. A qubit can be realized by quantum mechanical objects like a photon, where $\ket{0}$ and $\ket{1}$ states represent their polarization along two different axes. It can also be an electron at the ground ($\ket{0}$) state and excited ($\ket{1}$) state. In other words, the state of a qubit is a vector in a two-dimensional complex vector plane, and $\ket{0}$ and $\ket{1}$ form a set of orthonormal basis states~\cite{nielsen2001quantum}. Besides, qubits exhibit a property of entanglement, \ie, if two qubits are entangled, measuring one qubit reveals the state of the other. These properties give quantum computers an edge over classical computers.
More information on the fundamentals of quantum computing can be found in~\cite{8681202}.

\subsection{\textbf{General Framework of Quantum Machine Learning Workflow for Anomaly Detection}}
Figure~\ref{fig:QML_framework} shows a general workflow of QML processes distinguished by classical and quantum parts. 
\begin{itemize}
    \item Initially, we clean and standardize the data and run dimensionality reduction techniques like LDA (Linear Discriminant Analysis), PCA (Principle Component Analysis), or t-SNE (t-distributed Stochastic Neighbour Embedding) to fit the data in the near-term quantum models. Dimensionality reduction is essential as the existing quantum machines can handle smaller dimensional data due to the limited number of qubits.
    \item Next, encode the classical data into quantum states, which allows further quantum processing~\cite{nielsen2001quantum}. 
    \item The quantum states are processed by a quantum circuit~\cite{thapliyal2019design}, which maps the data into a higher dimensional feature space. The circuit is designed to maximize the amplitude of a specific output state that is the solution of the objective function.
    \item We can optimize the quantum circuit by repeated measurements and different gradient-based approaches using classical computers~\cite{schuld2020circuit}.
\end{itemize}

After training and deploying the model, we follow the same path to get predictions on new data. New data is first dimensionally reduced to match the input dimension of the model. Then, it is encoded into the quantum realm and processed by the quantum circuit. Ultimately, based on the measured expectation value, the model predicts if the data is an anomaly. Details of these individual steps of the quantum part are discussed in the next section.
\begin{figure*}[htb]
    \centering
    \includegraphics[width=2\columnwidth]{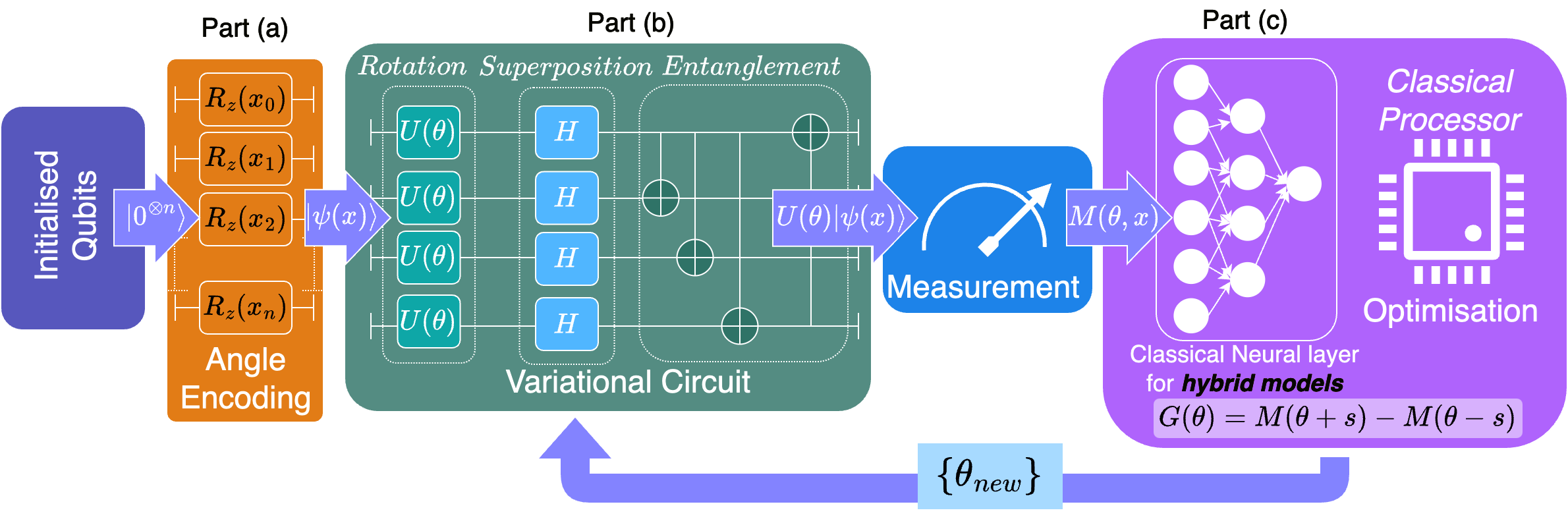}
    \caption{Variational Model for QML; Part(a): Quantum encoding, Part(b): Parameterised Quantum Circuit, Part(c): Classical Optimization}
    \label{fig:Variational_QML_Model}
\end{figure*}
\subsection{\textbf{QML Algorithms}}
QML circuit models can broadly be classified into variational circuits and kernel-based circuits.
\subsubsection{Variational Quantum Circuit (VQC)} Variational quantum machine learning algorithms use a Parameterized Quantum Circuit (PQC), a sequence of quantum gates controlled by tunable classical parameters. During training, we optimize the parameters of the PQC in a classical computer to minimize the cost function. Let us look into each component separately to realize how it works. The workflow of a VQC has primarily four steps:
\begin{itemize}
    \item In the first step, we encode the classical data into quantum states (We restrict it to only classical anomaly detection problems). Among many existing encoding techniques, angle encoding is the most popular for its simplicity, depicted in part (a) of Figure~\ref{fig:Variational_QML_Model}. Here, we apply unitary operations controlled by the data on a basis state (\eg, $\ket{0^{\otimes n}}$, for \textit{n}-dimensional classical data) to encode the classical information in the output quantum state.
    
    \item The second step, shown in Figure~\ref{fig:Variational_QML_Model}, part (b), is to send the encoded quantum states to a PQC, also known as Ansatz, which comprises parameterized quantum rotation gates, controlled by classical parameters and entangling layers consisting of Hadamard and Controlled NOT gate. The ideal design of a variational quantum model is still an open question. Expressivity is a good metric, referring to the sampling power of the circuit from a high-dimensional \textit{Hilbert Space}.

    \item The third step is to measure the output state to get its expected value. Based on this numerical result, we draw inferences from the data. For example, we can do parity post-processing, measuring the first two qubits. Repeated measurement will give us a probability distribution over the four output states: $\ket{00}, \ket{01}, \ket{10}, \ket{11}$. We can make a binary decision after adding the expectation values for even and odd parity~\cite{havlivcek2019supervised}. For example, if the result has odd parity, the data is an anomaly; otherwise, it is normal.

    \item The last step is optimization, shown in Figure~\ref{fig:Variational_QML_Model}, part (c). In a classical computer, based on the extracted labels, we estimate the cost function of the algorithm and use optimizers like Gradient descent to backtrack and update the trainable parameters in the PQC. A simple optimization strategy can be calculating the gradient for any parameter by shifting the parameter by a small amount $s$ and measuring the output, then shifting it down by $s$ and measuring again, and finally subtracting the results~\cite{schuld2020circuit}. The gradient calculation can be realized by the following equation, where $G(\theta)$ refers to the gradient concerning any parameter $\theta$, $\braket{0|\psi(\theta)|0}$ refers to the measured expectation value in $\ket{0}$ basis. $G(\theta) = \braket{0|\psi(\theta+s)|0}-\braket{0|\psi(\theta-s)|0}$. 
\end{itemize}

\subsubsection{Quantum Kernels}
In contrast to the variational method, kernel methods show more similarity to the working principles of quantum models. One of the key ideas of quantum computing is that quantum mechanical principles allow us to analyze data in a very high-dimensional \textit{Hilbert space}, which can be accessed only by inner products and measurement. In~\cite{schuld2021supervised}, the authors showed that we can replace many near-term and fault-tolerant quantum models that can be realized through a general support vector machine with a kernel that calculates the distance between encoded quantum states. They emphasized that instead of Dirac vectors $\ket{\psi(x)}$, we can define the feature vectors as the density matrices, $\rho(x)=\ket{\psi(x)}\bra{\psi(x)}$, and in this \textit{linear} feature space, the most suitable measurement strategy can define a decision boundary. The idea boils down to finding the best measurement strategy that optimizes the objective function of the given machine-learning problem. We can map the data into a higher dimensional space using a quantum kernel, where the anomalies will be linearly distinguishable from the standard data.

The QML algorithms are classified into Supervised, Unsupervised, and Reinforcement learning.

\subsection{\textbf{Supervised QML}}
In a supervised QML classifier, we feed the labeled encoded classical data into a quantum circuit that outputs the probability distribution over different classes. After every iteration, we measure the output to get the probability distribution over different states. With post-processing, we bring relevant numerical values for anomaly detection. 
Next, we shall look into a few supervised QML algorithms.
\subsubsection{Quantum Neural Network (QNN)}
A Quantum Neural Network (QNN) is generally a variational circuit consisting of classically trainable parameters, which are to be optimized over the training iterations. It can learn any circuit that minimizes the cost during training. At the end of every iteration, we perform a measurement that gives a deterministic probability value of the data being an anomaly. To increase the expressivity of the circuit, we can repeat a fixed quantum layer or combination of layers over and over.

\subsubsection{Hybrid Quantum Classical Neural Network (HQCNN)}

The idea of a QNN can be extended into a Hybrid Quantum-Classical Neural Network (HQCNN). In a hybrid model, the classical data can first be processed by a classical neural network, reducing its feature dimension. Then, we process the extracted features by the quantum circuit. However, we can first reduce the feature dimensions classically, process through a quantum circuit, and then run a classical network, taking input from measuring the output of the quantum circuit. This integration is shown in part (c) of Figure~\ref{fig:Variational_QML_Model}, which performs more stably than a stand-alone PQC. More details on hybrid quantum neural networks can be found in~\cite{HQNN_Arthur}. Anomaly detection systems can benefit from that because classical models got bulky with the increase in novelty in anomaly types, and they should be more flexible.

\subsubsection{Quantum Support Vector Machine (QSVM)}
The support vector machine is a linear model. It can discriminate between data with a hyperplane. SVM uses a kernel trick for linearly inseparable data to map them into a higher-dimensional feature space to make them linearly separable. The kernel is a matrix containing all the inner products of the feature vectors, $(\kappa)_{p,k} = \vec{v_p}.\vec{v_k}$ for $ p,k\in[1, N]$, $N$ is the number of data points. According to Rebentrost~\etal~\cite{rebentrost2014quantumSVM}, quantum computers can evaluate the inner products faster, meaning faster evaluation of the kernel matrix. This property enhances the speed of operation for the kernel methods like QSVM. The authors proved that given a $2^n$ basis vector space $\ket{\chi}$, the training vectors $\vec{v_p}$ can be represented as a superposition of the basis states, $\ket{v^p} = \sum_{i=1}^{2^n}{\alpha_i \ket{\chi^i}}$, the kernel can be calculated by taking the partial trace of the corresponding density matrix $\ket{\chi}\bra{\chi}$ for the states $\ket{\chi_i}$ using equation ~\ref{Q-kernel}, where $N_x$ is the squared sum of the magnitudes of the basis vectors.
\begin{equation}\label{Q-kernel}
    tr_x[\ket{\chi}\bra{\chi}] = \frac{1}{N_x}\sum_{i=1}^{2^n}{\braket{\chi^j|\chi^i}|\vec{\chi^j}||\vec{\chi^i}|\ket{j}\bra{i}}
\end{equation}
In many problems, anomalies exhibit high similarity with regular data. As an attacker will always try not to get caught, he will ensure the anomalous traffic or malware looks very similar to normal data. In such cases, a QSVM model that can access a high-dimensional vector space to map data can distinguish between anomalies and regular data.

\subsection{Unsupervised QML}
Unsupervised algorithms learn from unlabelled data; they are used to draw insights from the data without any prior knowledge. We cannot access labeled data in many anomaly detection cases, where unsupervised approaches like clustering are helpful. Some popular clustering techniques include K-means, $\delta$-k-means, and agglomerative clustering. 

\subsubsection{Quantum clustering}
The core idea of clustering is calculating the minimum distance between a data point and the centroids, in other words, finding which centroid is most similar to the data point. In~\cite{lloyd2013quantum}, Lloyd~\etal~proposed a method to calculate the distance between two real, \textit{n}-dimensional vectors $\vec a$ and $\vec{b}$ with the help of a \textit{swap}-test. They prepared two states, $\ket{\psi}= \frac{1}{\sqrt{2}}(\ket{0,a}+\ket{1,b})$ and $\ket{\phi} = \frac{1}{\sqrt{Z}}(|\vec{a}|\ket{0}-|\vec{b}|\ket{1})$, where $Z = |\vec{a}|^2+|\vec{b}|^2$, and evaluated $|\braket{\psi | \phi}|^2$ with the help of swap test. Preparing these states is easy if we encode the length of a vector $\vec x$, as the inner product of its quantum state with itself $\braket{x|x} = |\vec x|^{-1/2}|\vec x|$. For this definition $|\vec a- \vec b|^2 = Z|\braket{\psi|\phi}|^2$ is true. In this proposed method, considering $\vec a$, as a feature vector and $\vec b$ as an adjacent centroid, we can calculate the nearest centroid, which is the main theme of clustering. In many anomaly detection problems, people like to visualize the data in lower dimensions if there are any visible clusters and outliers. In such cases, the outliers usually refer to the anomalies. Using quantum-clustering techniques, higher dimensional clustering makes it possible to differentiate anomalies potentially.

\subsubsection{Quantum ensemble methods}
Generally, the objective of the ensemble method is to build a model based on the prediction of multiple base models, reduce dependency on a single model, and reduce prediction error. One example of the ensemble is `Bagging', where the decision is made based on majority voting. In~\cite{macaluso2020quantum}, Antonio~\etal, proposed a quantum algorithm to implement an ensemble approach using bagging. The benefit of the quantum ensemble algorithm comes from superposition, as we can generate a superposition of multiple transformations of training data and process them parallelly. In anomaly detection, anomalous data is rare compared to the regular data in the dataset. In a skewed dataset comes the problem of overfitting and bias. As the quantum ensemble method, besides its added quantum advantages, uses a majority vote from multiple learners, these problems can be mostly avoided.

\subsection{Quantum Reinforcement Learning (QRL)}
Like classical reinforcement learning, quantum reinforcement learning has three subelements: a policy, a reward function, and an environment model bound to the agent. However, the QRL algorithms differ from the traditional Reinforcement Learning (RL) algorithms in policy, representation, parallelism, and state update operation.

In~\cite{dong2008quantum}, Dong~\etal~proposed a novel QRL method. They introduced a value-updating framework inspired by quantum-mechanical properties like superposition and parallelism. They used multiple qubit systems to represent states and actions that could be expanded in terms of an orthogonal set of eigenstates $\ket{S_n}$, or eigenactions $\ket{A_n}$. The QRL agent has to learn a policy to maximize the expected sum of the discounted rewards of each state. It uses the collapse postulate, \ie, the action will collapse to one of the eigen actions upon measurement, to make its action-selection policy. A specific unitary transformation can parallelly process all the states according to the temporal difference (TD) rule. Probability-amplitude updating is the key to experiencing `train-and-error' to learn better, which occurs based on the Grover iteration~\cite{grover1997quantum}.

QRL can be used to build reliable ADS~\cite{rajawat2023quantum}, as a QRL-based ADS can constantly learn from new experiences with a reward and penalty system.

\section{Case Studies on QML-based Approaches for Anomaly Detection}
The general idea of QML workflow can be easily applied to solve anomaly detection tasks. This section discusses five works on anomaly detection in the consumer electronics domain, which used quantum machine learning as a solution.

In~\cite{rajawat2023quantum}, Rajawat~\etal~focussed on assessing the security and vulnerabilities of IoMT (Internet of Medical Things) systems. To address this problem, they have proposed a fused semisupervised reinforcement learning model and compared the results against state-of-the-art classical and quantum machine learning approaches. They used quantum methods to clean the data and extract features fed to the QML models to train them to identify weaknesses and dangers of the IoMT infrastructure. There is a feedback system that refines and updates the models based on incoming new data and threat patterns. After adequate training, the system can quickly predict and classify the vulnerabilities and take the proper action. They used five IoMT devices (viz., heart rate monitor, insulin pump, smart inhaler, pacemaker, and fitness tracker) to build the experimental setup to showcase the performance of the proposed model. They all have different security threats, like Eavesdropping, DoS, and Data Tampering, with High, Medium, or Low vulnerabilities. The results are compared against classical reinforcement learning agents, Deep Neural Networks (DNN), Convolutional Neural Networks (CNN), and quantum DNN. The quantum deep learning-based models achieved an accuracy ranging from 92.23\% to 99.34\%. The authors concluded that quantum computing makes it possible to rapidly process big data from IoMT devices, which QML models leverage to reduce the computation time for intricate pattern recognition significantly.

\begin{table*}[htbp]
\caption{Recent Works on Anomaly/ Intrusion Detection using Quantum Machine Learning Algorithms}
    \centering
    \resizebox{2.1\columnwidth}{!}{
    \setlength{\tabcolsep}{2pt}
    \begin{tabular}{lccccccc}
    \toprule
    \makecell[c]{\bf Authors \&\\ \bf Year} & \bf Data & \bf Domain & \bf QML Algorithm   & \makecell[c]{\bf Feature \\ \bf Selection \& \\ \bf Redution} & \bf Device & \makecell[c]{\bf Metrics (best)}\\
    \midrule
    \makecell[c]{Rajawat~\etal~\cite{rajawat2023quantum}\\2023} & \makecell[l]{HRM \\Insulin Pump \\Smart Inhaler \\Pacemaker \\Fitness tracker data} & \makecell[c]{IoMT} & \makecell[c]{Fused Q-DNN Semi-\\Supervised Learning} & \makecell[c]{QFS} & \makecell[c] {IBM quantum \\computer} & \makecell[l]{Acc: $99.34\%$} \\ 
    \midrule

    \makecell[c]{Wang, Huang~\etal~\cite{QHDNN_DeepAD_Wang}\\2023} & \makecell[l]{MNIST \\Fashion MNIST } & \makecell[c]{Deep Image AD} & \makecell[c]{Hybrid Q-DNN} & \makecell[c]{ N.A. } & \makecell[c] {Rigetti's quantum \\SDK Forest} & \makecell[l]{AUC: $89.41\%$\\(MNIST) \\ AUC: $88.24\%$\\(F-MNIST) } \\ 
    \midrule

    \makecell[c]{Alona Sakhnenko~\etal~\cite{sakhnenko2022hybrid}\\2022} & \makecell[l]{Power plant dataset \\ Musk dataset \\ Arrhythmia dataset\\ Satlog dataset} & \makecell[c]{General AD} & \makecell[c]{HAE} & \makecell[c]{ Classical \\Encoding } & \makecell[c] {Qiskit\\ State Vector \\Simulator} & \makecell[l]{recall: $64.1\%$\\(Gas turbine) \\ $100\%$(Musk) \\ $72.7\%$(Arrhythmia) \\$76\%$(Satlog) \\  } \\ 
    \midrule

    \makecell[c]{Javaria~\etal~\cite{AMIN2023104710}\\2023} & \makecell[l]{Crime-UCF\\ Crime-UNI} & \makecell[c]{Surveillance\\ Anomaly detection} & \makecell[c]{J-QCNN} & \makecell[c]{ None } & \makecell[c]{Classical Simulator} & \makecell[l] {Acc: 100\% \\(Highest)}  \\ 
    \midrule

    \makecell[c]{Oleksandr, Einar~\cite{kyriienko2022unsupervisedFD}\\2022}& \makecell[l]{Credit Card\\ Fraud Detection \\dataset} & \makecell[c]{Finance} & \makecell[c]{Q-SVC\\Q-OC-SVM} & \makecell[c]{ PCA } & \makecell[c]{Classical Simulator,\\NVIDIA A100 GPU} & \makecell[c] {Average Precision: \\70\%}  \\ 
    \midrule
    \bottomrule
    \end{tabular}
    }
    \footnotesize{\break \\ HRM: Heart Rate Monitor, IoMT: Internet of Medical Things, QFS: Quantum Feature Selection, HAE: Hybrid Quantum Classical Auto Encoder, J-QCNN: Javeria Quantum and Convolutional Neural Network, Q-SVC: Quantum Support Vector Classifier, Q-OC-SVM: Quantum One-Class Support Vector Machine}
    \label{tab:comp}
\end{table*}

In~\cite{QHDNN_DeepAD_Wang}, Wang~\etal~proposed a Quantum Hybrid Deep Neural Network (QHDNN) model based on a classical anomaly detection (AD) technique, Deep Support Vector Data Description (DSVDD) for deep image AD. After extensive experiments, they concluded that their proposed model produced better accuracy while sharing the same number of learnable parameters. DSVDD employs a classical DNN to map raw features into a hypersphere whose objective is to minimize its volume. After training, the model maps the normal data inside the sphere, keeping the anomalies outside. The proposed QHDNN model consists of classical convolution and pooling layers in the beginning for feature extraction; later, the extracted features are fed to the VQC-based quantum layers (replacing fully connected layers) for feature mapping (into a hypersphere, inspired by DSVDD). They have used MNIST and Fashion-MNIST datasets for experimental analysis. Their idea was to train the model, considering any of the ten classes (\eg, x$\in[0,9]$) as standard and the rest as anomalies. Therefore, MNIST, a dataset of ten classes, allowed them to train it ten times, selecting any classes as normal against nine other classes as anomalies. They evaluated the performance in terms of the area under the receiver operating characteristic curve, AUC. The best performance in terms of AUC for Fashion-MNIST was recorded as 89.41\%, and MNIST was 88.24\%.

In~\cite{sakhnenko2022hybrid}, Sakhnenko~\etal~introduced a novel hybrid Quantum Classical Auto Encoder (HAE) for anomaly detection using near-term quantum computers. The HAE consists of a PQC between the classical encoder and decoder circuits. The anomalous data will produce unexpectedly high reconstruction loss in Auto Encoders (AE) trained with standard data. If the anomalies are embedded at unexpected places in the latent space of AE, they extract the latent space to apply an isolation forest and flag the anomalies more reliably. To test this approach in a real-world scenario, they chose a power plant dataset consisting of 161 sensor data monitoring the gas turbine, recorded every 10 minutes through 10 months. Due to computational complexity, they train the HAE using 640 data points(1 week's observation). The original data is compressed by the encoder and fed into a 4-qubit PQC. After measuring the expectation values of the qubits, they get the latent space of the HAE, which is used to train an isolation forest model. After testing on 2000 datapoints, the model yields 64.1\% recall. They benchmarked their model with publicly available Musk, Arrhythmia, and Satelite datasets, showing significant performance increases over classical AE in precision, recall, and F1 score. They have also tested 30 different PQC designs to choose the best one for HAE, concluding that entanglement significantly impacts model performance.

In~\cite{AMIN2023104710}, Javaria~\etal~discussed QML methods for detecting anomalies, mostly violent crimes like armed thefts and robberies, in a sequence of video frames using Quantum Convolutional Neural Networks (QCNNs). The Javeria-QCNN ( J-QCNN) model consists of initial dense layers, followed by drop-out, flattening, and final dense layers that indicate whether the frame is anomalous or normal. They chose the Crime-UCF dataset for the experiment, containing 1900 video surveillance of regular events, crime, robbery, arson, vandalism, and other anomalies. They also used the Crime-UNI dataset, created based on Crime-UCF, consisting of 12,810 frames of robbery and 30,030 total frames of standard video footage. The lengths of these videos are standardized by using only the practical 10 seconds. They compared their model against a classical Javeria Deep Convolutional Neural Network (J-DCNN) model on the same grounds. In different classification scenarios, J-DCNN showed accuracies such as 0.97 (robbery/ normal), 0.93 (abuse/ arrest/ arson/ normal), 0.94 (burglary/ explosion/ fighting/ normal), and 0.90 (stealing/shooting/shoplifting). In contrast, J-QCNN yielded 0.98  (robbery/ normal), 0.99 (abuse/ arrest/ arson/ normal), 1.00 (burglary/ explosion/ fighting/ normal), and 0.93 (stealing/ shooting/ shoplifting). From the results, it is clear that QCNN outperforms the classical DCNN.

In~\cite{kyriienko2022unsupervisedFD}, Oleksandr~\etal~developed a fraud detection model inspired by quantum kernels using the credit card dataset. Initially, they use PCA and take the first $N$ components. $N$ is the number of qubits, which varies from a few to 20. They encode this feature space using an instantaneous quantum polynomial (IQP) circuit, which maps the data into a higher dimensional space, enabling the kernel trick. They explored multiple ways to construct the Gram matrix, populated by the distances between individual data vectors. Once this kernel is built, the rest of the workflow follows a one-class support vector machine. They have constructed the IQP circuit with data re-uploaded three times to increase the expressivity of their model. The authors tested their model against unsupervised kernel-based benchmark methods, and they established that the quantum kernel-based classifier outperforms the classical kernel by 15\% in average precision in an increased feature space, which can save millions of pounds from fraud once implemented. They also discussed the practical aspects of their approach. The main idea of any kernel-based classifier is calculating the Gram matrix, which is an $O(N_s^2)$ order operation, $N_s$ being the number of data points. This step takes a long time with the increased data volume.
On top of that, for any new data, an operation of the order of $O(N_s)$ will be required to re-evaluate the kernel. Also, for inferencing $N_d$ number of data, the computational cost will be an order of $O(N_dN_s)$, which increases the system's reaction time. This time complexity led them to conclude that quantum kernel-based methods can be applied for only re-checking transactions and evaluating top-priority cases instead of mass evaluations in real time.

\section{Discussion}
Table~\ref{tab:comp} summarises the case studies in this work. The case studies show the relevance of quantum machine learning in complex anomaly detection tasks, which are currently solved by classical methods at the industrial scale. Most researchers have used a hybrid approach, yielding better results than the pure quantum approach. Fitting raw data directly into the quantum circuit is challenging due to the limitation of the total number of usable qubits. Therefore, in most cases, raw data is dimensionally reduced before fitting into a quantum circuit. In some scenarios (\eg,~\cite{sakhnenko2022hybrid}), data is initially fed to a classical network that reduces its dimension before quantum processing. The quantum circuits can not handle large volumes of data, so we saw a trend of subsampling the original dataset before application. Irrespective of these limitations, a quantum model with comparable parameters can perform better than a classical network~\cite{QHDNN_DeepAD_Wang}. The key takeaways from this work are as follows:
\begin{itemize}
    \item Anomaly detection is a potential field of application for quantum machine learning for its flexibility and the novelty of the approach.
    \item Many supervised, unsupervised, and reinforcement learning QML algorithms have been researched and found to apply to anomaly detection problems effectively.
    \item Though QML algorithms are still in their infancy, the future emergence of fault-tolerant quantum computers will realize the industrial applications of QML algorithms in building ADS.
    \item In some cases, theoretically, quantum algorithms yield higher performance than their classical counterparts, such as J-QCNN~\cite{AMIN2023104710}.
    \item Mapping classical data into a complex higher dimensional quantum domain helps solve many anomaly detection problems.
\end{itemize}

\section{Conclusion and Future Work}
This paper has discussed how we can leverage the properties of quantum machine learning to solve anomaly detection problems. In the rapidly changing world, bulky classical models are inefficient in providing flexibility, whereas quantum networks are light and flexible. Despite quantum computers being noisy and less reliable, researchers are trying to figure out new algorithms to solve machine-learning tasks in near-term quantum computers. Near-term quantum algorithms flourished after the innovation of Variational Quantum Eigensolver~\cite{tilly2022VQE}. Many other quantum-enhanced processes have been introduced to speed up machine-learning tasks~\cite{biamonte2017QML}. However, the error encountered in a near-term quantum device imposes significant uncertainties during operation. One of the major practical problems in quantum computers is decoherence~\cite{shor1995scheme}. As it is impossible to isolate a quantum system entirely, its information decays and gives rise to errors. Though a practical quantum computer will probably be devised soon, quantum error correction will be needed to envision a fault-tolerant quantum computer~\cite{preskill1998fault}. 

However, many studies show the relevance of the quantum version of neural networks, one of the most primitive and practical building blocks of artificial intelligence. Unlocking the quantum world's full potential can open new opportunities for future scientists. Therefore, we expect significant research advancement in various consumer electronics domains, including anomaly detection with the help of quantum machine learning. 

\balance

\bibliographystyle{IEEEtran}
\bibliography{IEEEabrv}

\begin{thebibliography}{10}
\providecommand{\url}[1]{#1}
\csname url@samestyle\endcsname
\providecommand{\newblock}{\relax}
\providecommand{\bibinfo}[2]{#2}
\providecommand{\BIBentrySTDinterwordspacing}{\spaceskip=0pt\relax}
\providecommand{\BIBentryALTinterwordstretchfactor}{4}
\providecommand{\BIBentryALTinterwordspacing}{\spaceskip=\fontdimen2\font plus
\BIBentryALTinterwordstretchfactor\fontdimen3\font minus \fontdimen4\font\relax}
\providecommand{\BIBforeignlanguage}[2]{{%
\expandafter\ifx\csname l@#1\endcsname\relax
\typeout{** WARNING: IEEEtran.bst: No hyphenation pattern has been}%
\typeout{** loaded for the language `#1'. Using the pattern for}%
\typeout{** the default language instead.}%
\else
\language=\csname l@#1\endcsname
\fi
#2}}
\providecommand{\BIBdecl}{\relax}
\BIBdecl

\bibitem{10398426}
R.~Pell, S.~Moschoyainnis, and M.~Shojafar, ``{LSTM} based anomaly detection of pfcp signaling attacks in 5g networks,'' \emph{IEEE Consumer Electronics Magazine}, pp. 1--9, 2024.

\bibitem{maheshwari2022quantum}
D.~Maheshwari, B.~Garcia-Zapirain, and D.~Sierra-Sosa, ``Quantum machine learning applications in the biomedical domain: A systematic review,'' \emph{IEEE Access}, vol.~10, pp. 80\,463--80\,484, 2022.

\bibitem{nielsen2001quantum}
M.~A. Nielsen and I.~L. Chuang, \emph{Quantum computation and quantum information}.\hskip 1em plus 0.5em minus 0.4em\relax Cambridge university press Cambridge, 2001, vol.~2.

\bibitem{8681202}
T.~S. Humble, H.~Thapliyal, E.~Muñoz-Coreas, F.~A. Mohiyaddin, and R.~S. Bennink, ``Quantum computing circuits and devices,'' \emph{IEEE Design \& Test}, vol.~36, no.~3, pp. 69--94, 2019.

\bibitem{thapliyal2019design}
H.~Thapliyal and E.~Mu{\~n}oz-Coreas, ``Design of quantum computing circuits,'' \emph{IT Professional}, vol.~21, no.~6, pp. 22--26, 2019.

\bibitem{schuld2020circuit}
M.~Schuld, A.~Bocharov, K.~M. Svore, and N.~Wiebe, ``Circuit-centric quantum classifiers,'' \emph{Physical Review A}, vol. 101, no.~3, p. 032308, 2020.

\bibitem{havlivcek2019supervised}
V.~Havl{\'\i}{\v{c}}ek, A.~D. C{\'o}rcoles, K.~Temme, A.~W. Harrow, A.~Kandala, J.~M. Chow, and J.~M. Gambetta, ``Supervised learning with quantum-enhanced feature spaces,'' \emph{Nature}, vol. 567, no. 7747, pp. 209--212, 2019.

\bibitem{schuld2021supervised}
M.~Schuld, ``Supervised quantum machine learning models are kernel methods,'' 2021.

\bibitem{HQNN_Arthur}
D.~Arthur and P.~Date, ``Hybrid quantum-classical neural networks,'' in \emph{2022 IEEE International Conference on Quantum Computing and Engineering (QCE)}, 2022, pp. 49--55.

\bibitem{rebentrost2014quantumSVM}
P.~Rebentrost, M.~Mohseni, and S.~Lloyd, ``Quantum support vector machine for big data classification,'' \emph{Physical review letters}, vol. 113, no.~13, p. 130503, 2014.

\bibitem{lloyd2013quantum}
S.~Lloyd, M.~Mohseni, and P.~Rebentrost, ``Quantum algorithms for supervised and unsupervised machine learning,'' 2013.

\bibitem{macaluso2020quantum}
A.~Macaluso, S.~Lodi, C.~Sartori \emph{et~al.}, ``Quantum algorithm for ensemble learning.'' in \emph{ICTCS}, 2020, pp. 149--154.

\bibitem{dong2008quantum}
D.~Dong, C.~Chen, H.~Li, and T.-J. Tarn, ``Quantum reinforcement learning,'' \emph{IEEE Transactions on Systems, Man, and Cybernetics, Part B (Cybernetics)}, vol.~38, no.~5, pp. 1207--1220, 2008.

\bibitem{grover1997quantum}
L.~K. Grover, ``Quantum mechanics helps in searching for a needle in a haystack,'' \emph{Physical review letters}, vol.~79, no.~2, p. 325, 1997.

\bibitem{rajawat2023quantum}
A.~S. Rajawat, S.~Goyal, P.~Bedi, T.~Jan, M.~Whaiduzzaman, and M.~Prasad, ``Quantum machine learning for security assessment in the internet of medical things (iomt),'' \emph{Future Internet}, vol.~15, no.~8, p. 271, 2023.

\bibitem{QHDNN_DeepAD_Wang}
M.~Wang, A.~Huang, Y.~Liu, X.~Yi, J.~Wu, and S.~Wang, ``A quantum-classical hybrid solution for deep anomaly detection,'' \emph{Entropy}, vol.~25, no.~3, 2023.

\bibitem{sakhnenko2022hybrid}
A.~Sakhnenko, C.~O’Meara, K.~J. Ghosh, C.~B. Mendl, G.~Cortiana, and J.~Bernab{\'e}-Moreno, ``Hybrid classical-quantum autoencoder for anomaly detection,'' \emph{Quantum Machine Intelligence}, vol.~4, no.~2, p.~27, 2022.

\bibitem{AMIN2023104710}
J.~Amin, M.~A. Anjum, K.~Ibrar, M.~Sharif, S.~Kadry, and R.~G. Crespo, ``Detection of anomaly in surveillance videos using quantum convolutional neural networks,'' \emph{Image and Vision Computing}, vol. 135, p. 104710, 2023.

\bibitem{kyriienko2022unsupervisedFD}
O.~Kyriienko and E.~B. Magnusson, ``Unsupervised quantum machine learning for fraud detection,'' 2022.

\bibitem{tilly2022VQE}
J.~Tilly, H.~Chen, S.~Cao, D.~Picozzi, K.~Setia, Y.~Li, E.~Grant, L.~Wossnig, I.~Rungger, G.~H. Booth \emph{et~al.}, ``The variational quantum eigensolver: a review of methods and best practices,'' \emph{Physics Reports}, vol. 986, pp. 1--128, 2022.

\bibitem{biamonte2017QML}
J.~Biamonte, P.~Wittek, N.~Pancotti, P.~Rebentrost, N.~Wiebe, and S.~Lloyd, ``Quantum machine learning,'' \emph{Nature}, vol. 549, no. 7671, pp. 195--202, 2017.

\bibitem{shor1995scheme}
P.~W. Shor, ``Scheme for reducing decoherence in quantum computer memory,'' \emph{Physical review A}, vol.~52, no.~4, p. R2493, 1995.

\bibitem{preskill1998fault}
J.~Preskill, ``Fault-tolerant quantum computation,'' in \emph{Introduction to quantum computation and information}.\hskip 1em plus 0.5em minus 0.4em\relax World Scientific, 1998, pp. 213--269.

\end{thebibliography}

\end{document}